\documentclass[epj]{webofc}
\usepackage[varg]{txfonts}   
\usepackage{amsmath,amssymb}
\usepackage{graphicx}
\graphicspath{{./Figs/}}
\usepackage{floatflt}
\usepackage{marginnote}
\usepackage{color}
\wocname{EPJ Web of Conferences}
\woctitle{CONF12}
\newcommand{\Tr}{\operatorname{Tr}}
\renewcommand{\Re}{\mathfrak{Re}}
\renewcommand{\Im}{\mathfrak{Im}}

\newlength{\currentparskip}
\newlength{\currentparindent}
\begin{document}
\selectlanguage{english}
\title{QCD propagators and vertices from lattice QCD }
\subtitle{(in memory of Michael Müller-Preu\ss{}ker)}

\author{Andr\'e Sternbeck\inst{1}\fnsep\thanks{\email{andre.sternbeck@uni-jena.de}}}

\institute{Theoretisch-Physikalisches Institut, 
    Friedrich-Schiller-Universit\"at Jena, 07743 Jena, Germany}

\abstract{%
  We review lattice calculations of the elementary Greens functions of
QCD with a special emphasis on the Landau gauge. These lattice results
have been of interest to continuum approaches to QCD over the past 20
years. They are used as reference for Dyson-Schwinger- and functional
renormalization group equation calculations as well as for hadronic
bound state equations. The lattice provides low-energy data for
propagators and three-point vertices in Landau gauge at zero and
finite temperature even including dynamical fermions. We summarize Michael
M\"uller-Preu\ss{}ker's important contributions to this field and put them 
into the perspective of his other research interests. 
}
\maketitle
%
\section{Michael's scientific career and research topics}
\label{sec:intro}

Michael M{\"u}ller-Preu\ss{}ker died much too early on October 12, 2015, while
on a business trip to a lattice gauge theory workshop in Vladivostok, Russia. 
Although he had retired four years ago he still held a Senior Professorship at the 
physics institute of the Humboldt-University Berlin and, active as he was, 
continued with his research and teaching until the very end. He was a Professor 
of Theoretical Physics with his heart and soul and a well-appreciated colleague 
and teacher at his Alma mater.

\medskip
\hbox{\begin{minipage}[b]{0.75\textwidth} 
\setlength{\parindent}{\currentparindent}
Michael was born on September 26, 1946 in Potsdam and received the
\emph{``Abitur''}, the entrance degree for German universities, in 1965. 
He enrolled as physics student at Humboldt-University which awarded him 
the doctoral degree (\emph{Dr.\ rer.\ nat.}) in 1973 for his work on ``Sum Rules For Helicity 
Partial Wave Amplitudes'' under the supervision of Prof.\ F.\ Kaschluhn. He 
remained at the physics institute also for his postdoctoral studies, employed 
as a Research Assistant (``\emph{Assistent}'' and ``\emph{Oberassistent}'') in 
the research unit \emph{Particles and Fields} from 1972 to 1993. During that 
time, he was delegated to the Joint Institute for Nuclear Research (JINR)
where he spent 5 years as a Visiting Researcher from 1978 to 1983, a period 
which was very valuable for his scientific career. Ever since then he 
maintained close relations with his Russian colleagues and there are many 
joint publications. In 1986 he received the doctoral degree (\emph{Dr.\ sc.\ nat.})
for his work on instantons in Euclidean Yang-Mills theory on the lattice and continuum. 
\end{minipage}\;
\begin{minipage}[b]{0.25\textwidth}
\includegraphics[width=3.3cm]{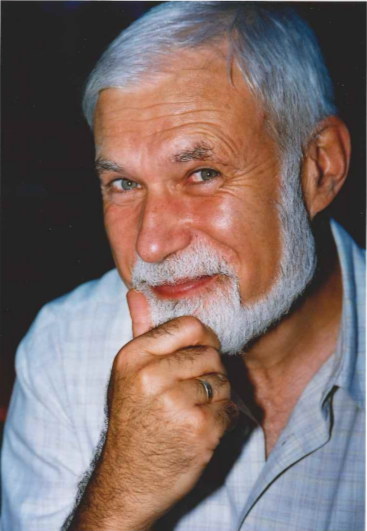}
\vspace{0.9cm}
\end{minipage}}

\noindent\begin{minipage}[b]{0.29\textwidth}
\includegraphics[width=3.95cm]{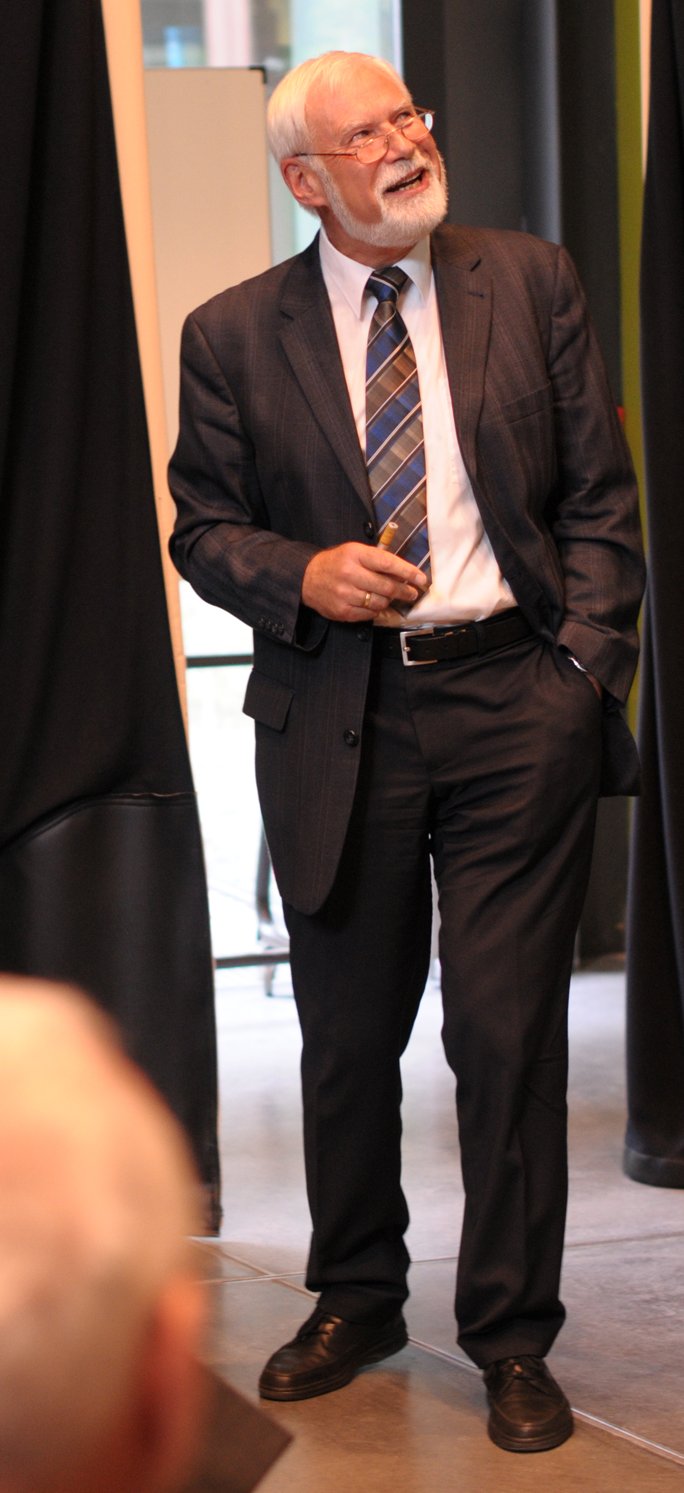}
\end{minipage}\;
\begin{minipage}[b]{0.7\textwidth}
\setlength{\parindent}{\currentparindent}
After the German reunification Michael was Visiting Professor at the University of 
Bielefeld (1990--1991) and then after a short period as Senior Scientist became 
Professor for Theoretical Physics at the Humboldt-University Berlin in 1993. Since then he 
educated quite a few physics students and postdocs in his research group on 
\emph{``Phenomenology / Lattice Gauge Theory''}. In fact, Michael was a passionate 
and enthusiastic teacher and introduced students to research at an early stage.
After his retirement Michael continued with research and teaching on the basis of 
a Senior Professorship in 2011.

Michael also had quite a few university responsibilities. To name but a few:
From 1994 to 1996 he was Vice-President of the Humboldt-University Berlin, a
period which was still characterized by dramatic changes in the entire 
social environment and it did not spare people at the university.
From 1997 to 2003 Michael was head of the task force ``Adlershof'' which 
organized the relocation of the Institute of Physics from the city center to 
the new science campus in Berlin-Adlershof. During that time he was also 
chairman of the Local Organizing Committee of the XIX International Symposium 
on Lattice Field Theory in 2001. From 2002 to 2006 he was Vice-Dean 
for Teaching and Studies at the Faculty for Math \& Science~I and
so took part in the introduction of the new Bachelor and 
Master degrees. Later he helped to reintroduce 
\end{minipage}
\noindent the annual  graduation 
ceremony at the institute, a tradition which had disappeared for years.
From 2006 to 2010 Michael was Managing Director of the Institute of Physics.
After his retirement he was Chairman of the Physical Society Berlin (PGzB) from 
2012 to 2014 and member of the scientific board of the North-German Supercomputing 
Alliance (HLRN). 

\medskip
Michael's primary research topics were (1) topology in Yang-Mills theories,
(2) gauge fixing, in particular the Gribov copy problem, and (3) QCD 
thermodynamics during the last years. 
He also looked at other topics like, for example, spin and gauge Higgs models or 
quasi Monte Carlo methods and many more. Michael always 
had a good overview about the different developments in the lattice community 
and from time to time started to work on something new.

Topology was indeed his main subject and his studies were often inspired by
semiclassical physics. He became interested in topology during his time in Dubna 
together with his long-term collaborator Ernst-Michael Ilgenfritz. Their first 
paper appeared in 1979 and 
was about the phase transition in the Yang-Mills instanton gas 
\cite{Ilgenfritz:1979nh}. Topology was also their motivation to start with lattice 
simulations. In the early 80's these simulations were just about
to gain momentum and their first lattice-based topology results appeared in 
1982. First numerical evidence for instantons in the vacuum of a SU(2) lattice 
gauge theory was then provided in 1986 \cite{Ilgenfritz:1985dz}, which was
an important contribution at that time.
\nocite{Bornyakov:1990es,Bornyakov:1991se,Bali:1996dm}

\medskip
\hbox{%
\begin{minipage}[b]{0.8\textwidth}
 \setlength{\parindent}{\currentparindent}
Many more studies followed, not only on instantons but also on monopoles 
(e.g., \cite{Bornyakov:1990es,Bornyakov:1991se,Bali:1996dm}) to address the 
confinement problem in the context of the maximal Abelian gauge and Abelian 
dominance (dual superconductor scenario by 't Hooft and Mandelstam), and later 
also on calorons and dyons. Often these were collaborative work together with 
Ilgenfritz and their Russian colleagues and long-term collaborators Mitrjushkin,
Martemyanov and Bornyakov, or later with van Baal, Bruckmann and Gattringer (e.g.,
\cite{Bruckmann:2004zy,Bruckmann:2006wf}). 
Michael's last topology paper was about dyons near the QCD transition temperature 
\cite{Bornyakov:2015xao} and appeared at the end of 2015.
\end{minipage}
\quad
\begin{minipage}[b]{0.2\textwidth}
 \includegraphics[width=2.3cm]{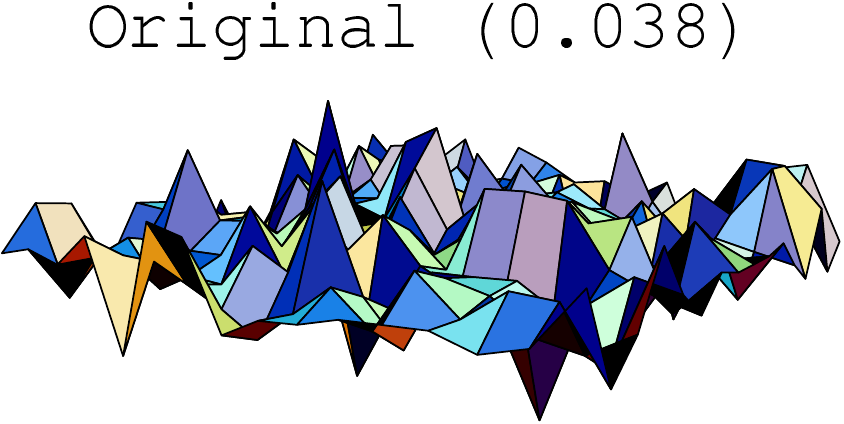}\\*[1.5ex]
 \includegraphics[width=2.3cm]{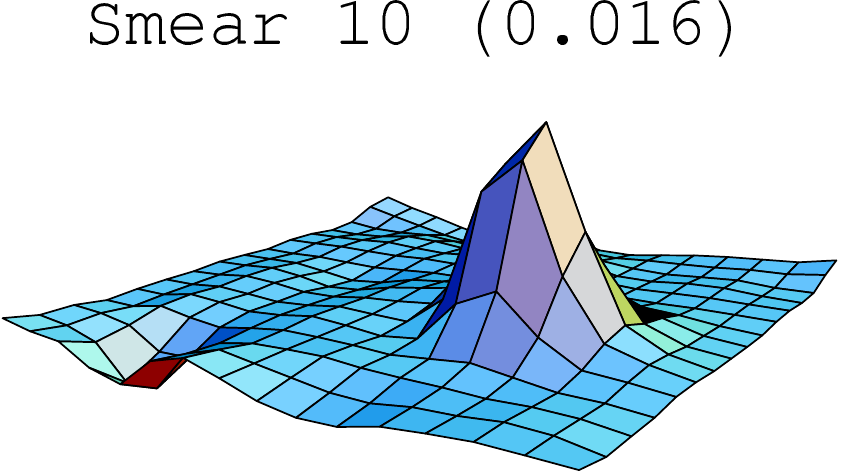}\\
{\footnotesize Figures taken from \cite{Bruckmann:2006wf}}
\end{minipage}
}

\clearpage

\begin{floatingfigure}[l]{0.43\textwidth}
 \hspace{-1.5em}
 \includegraphics[width=6cm]{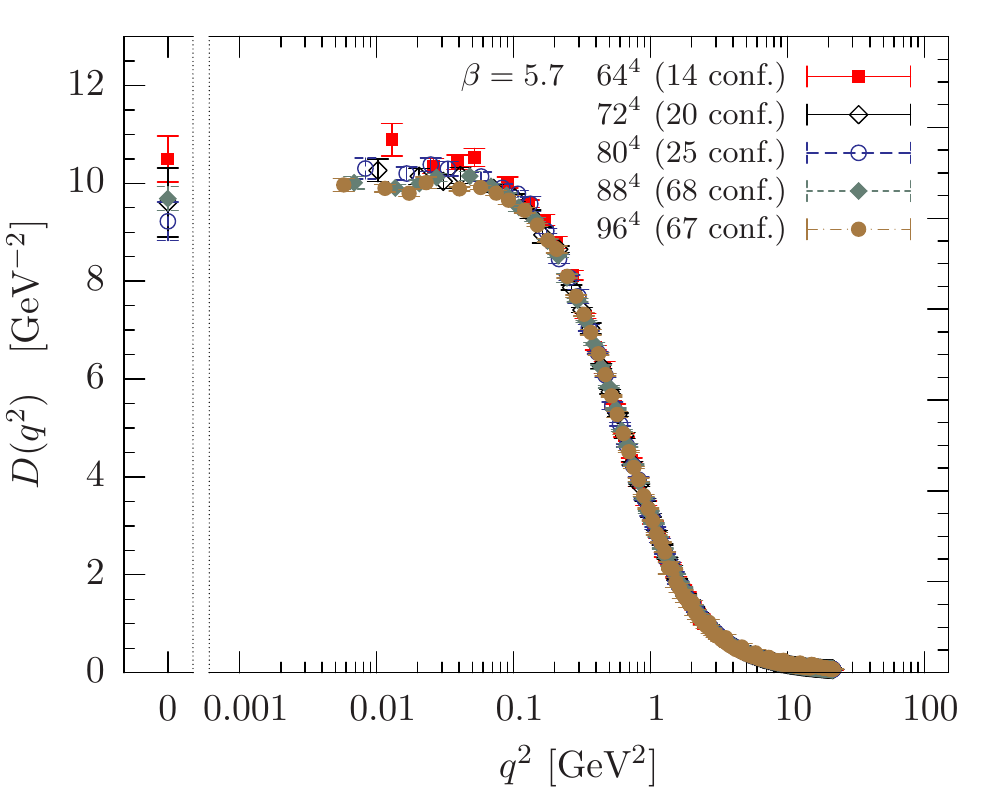}
 \caption{Gluon propagator vs.\ $q^2$; from \cite{Bogolubsky:2009dc}.}
 \label{fig:gl_qq_beta5p7}
\end{floatingfigure}
Michael's second main topic were lattice gauge field theories supplemented
with a gauge condition. In particular the problem of Gribov copies and its
impact on gauge-variant propagators or Abelian observables was something he 
worked on every now and then. His studies did not only focus on the SU(N) 
gluon and ghost propagators in Landau gauge, which indeed have received broad 
interest \cite{Bogolubsky:2009dc} more recently. Prior to this, Michael 
performed several lattice studies of compact QED in Lorentz (Landau) gauge 
which aimed at a nonperturbative formulation of QED on the lattice without 
magnetic monopoles (e.g., \cite{Bornyakov:1993yy}).
Many of these studies were performed together with his Russian colleagues 
Bogolubsky, Bornyakov and Mitrjushkin. These studies also led him to the 
Gribov problem, an ambiguity in the gauge condition which was addressed then in many
of his subsequent studies. On the lattice this problem is similarly present in the
Abelian projection and so as well as relevant for their lattice studies on the dual 
superconductor scenario. In \cite{Bali:1996dm}, Gunnar Bali together with 
Michael and others introduced therefore a new gauge-fixing (maximization) 
prescription which builds upon the method of simulated annealing to reduce the 
impact of the Gribov ambiguity. This prescription reappeared later in many of 
Michael's studies of SU(2) and SU(3) Yang-Mills theories in Landau or Coulomb 
gauge, for instance, in the lattice studies of the low-momentum behavior of 
the gluon and ghost propagators which fascinated him for some years. This 
interest was triggered around 2002 by new developments for the treatment of 
QCD with functional methods and hence a chance for interdisciplinary research 
arose. Many lattice studies on that followed and for their efforts the 
Russian-German group of authors, including Michael together with Bogolubsky, Bornyakov, 
Ilgenfritz, Mitrjushkin and myself, received in a JINR award in 2015 for their 
successful long-year collaboration. In recent years Michael took part in new 
investigations of the 3-gluon and quark-gluon vertex in Landau gauge, which is 
still ongoing.

\medskip
\begin{floatingfigure}[r]{0.42\textwidth}
 \vspace{1ex}
 \hspace{-1.5em}
 \includegraphics[width=6cm]{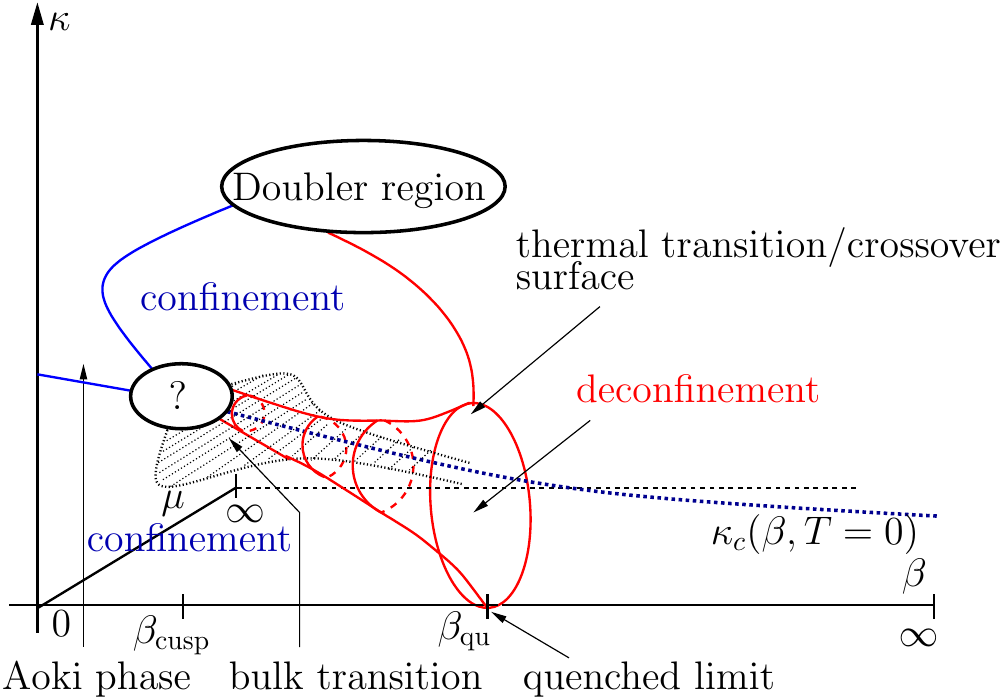}
 \caption{Schematic view of the phase diagram; Figure taken from \cite{Ilgenfritz:2009ns}.}
\end{floatingfigure}
Since the 90's Michael participated in various lattice calculations in 
QED or QCD with dynamical fermions. In particular, he was very enthusiastic
about QCD thermodynamics. In 2005 this interest became more intensive and 
focused. A few years before, first large-scale lattice QCD calculations with
twisted-mass fermions were performed by members of the nearby NIC/DESY group in 
Zeuthen to overcome problems of other lattice fermion actions. It was Michael's
idea to complement the efforts of the ETM collaboration by corresponding 
studies  at finite temperature, in particular as such studies with Wilson-type 
fermions were rare at that time.

During the first years, the \emph{tmfT Collaboration}
\footnote{Members of the tmfT Collaboration were: Burger, Ilgenfritz, Jansen, Kirchner, 
Lombardo, M\"uller-Preussker, Petschlies, Philipsen, Pinke and Zeidlewicz 
(though not all at the same time).} 
focused on the phase structure in the temperature-mass plane, i.e.,
the three-dimensional parameter space for temperature, bare 
quark mass (hopping parameter) and twisted-mass parameter $\mu$ \cite{Ilgenfritz:2009ns,Burger:2011zc}. 
Of special interest were the thermal phase transition and the location of
the Aoki phase \cite{Aoki:1983qi}.
\footnote{The Aoki phase for plain Wilson fermions at zero and finite temperature
 was also the subject of my diploma thesis in 2002 
 \cite{Ilgenfritz:2003gw,Ilgenfritz:2005ba}. Michael's supervision and support was 
 excellent and I am very grateful that he introduced me to lattice field theories.
}
These preparatory studies turned out to be quite involved due to the richer
phase structure in the three-dimensional parameter space (see figure above). 
Results for the thermal cross-over and for the equation of state appeared 
then at the end of 2014 with emphasis on the continuum limit because physical 
quark masses were not yet in reach \cite{Burger:2014xga}. This study was then 
continued with $N_f=2+1+1$ fermion flavors to quantify contributions from 
heavy quarks. In addition to that, the temperature dependence of the 
topological susceptibility was analyzed. First $N_f=2+1+1$ results appeared 
in a few conference proceedings \cite{Burger:2015xda, Trunin:2015yda} but this 
part has remained unfinished since then.
An almost up-to-date summary of the tmfT results was given by Florian Burger 
and Michael in \cite{Burger:2015era} to which I would like to 
refer the reader for further details.

During the last years Michael also looked at two-color QCD at finite 
temperature and analyzed with his students (and colleagues) in Berlin and 
Russia the influence of (1) a magnetic field and (2) that of a finite chiral chemical 
potential \cite{Braguta:2015zta}. For those studies dynamical staggered fermions were used instead of 
Wilson twisted-mass. When switching on a magnetic field they found numerical 
evidence for magnetic catalysis and inverse catalysis 
\cite{Ilgenfritz:2012fw,Ilgenfritz:2013ara} (at different temperatures) which 
was kind of a hot topic lately. 

\medskip
\centerline{----------------------------------------}

Michael gave an overview on the present status of his first main topic (topology)
at the lattice conference 2014 and devoted this to the memory of Pierre van
Baal \cite{Muller-Preussker:2015daa}. In my contribution I will focus on his 
second main topic and summarize our joint lattice studies (2003--2015). I 
apologize that I do not have the space to review all developments in this
field and concentrate on the lattice studies Michael was involved. 

\section{Gauge-variant Green's functions of QCD}

Lattice-QCD practitioners typically are rather skeptical about the usefulness 
analyzing gauge-variant quantities on the lattice. Gauge invariance at any finite 
lattice spacing is the distinguishing feature of the lattice formulation, so 
why bother worrying about a gauge, related quantities and problems?

If one is just interested in numbers for gauge-invariant quantities like 
hadron masses, form factors or the like, one indeed does not need to worry 
about a gauge. An additional condition on the gauge field would simply drop out 
in the vacuum expectation value of a gauge-invariant object.

The lattice, however, allows for more. It provides nonperturbative access to 
the elementary \hbox{$n$-point} Green's functions of QCD whose functional 
dependence on momentum or distance is intimately connected with a gauge. The 
quark, gluon and ghost 2-point functions (propagators) in Landau gauge are ever 
popular examples. Also 3-point functions, like the 3-gluon, quark-gluon or 
ghost-gluon vertices or any higher $n$-point vertex function are accessible. 
The signal-to-noise ratio though decreases the larger $n$. 

Still, why bother about gauge-variant $n$-point functions? At present I can 
provide two answers:

\begin{enumerate}
 \item Lattice calculations have demonstrated great potential for the study 
       of hadronic properties and of QCD in general. 
       Beside the lattice there are however also other nonperturbative 
       frameworks which allow to address strong interaction physics. 
       One example are the flow equation of the Functional Renormalization Group
       (FRG, aka Wetterich equation \cite{Wetterich:1992yh}), another 
       the bound-state and Dyson-Schwinger equations (DSEs) of QCD.\footnote{The reader 
       may consult the reviews \cite{Roberts:1994dr,Alkofer:2000wg,Gies:2006wv,Eichmann:2016yit} 
       and the references therein.} As the lattice, both these
       approaches start from first principles, namely the effective average 
       action and the partition function, respectively, but in addition they
       require a gauge. Any physics extracted with either method is independent 
       of the gauge, for example, masses or form 
       factors via correspondingly defined bound-state equations. In practice 
       though, the numerical treatment requires a truncation, either of the 
       infinite tower of DSEs or of the infinite hierarchy of integro-differential 
       (FRG) equations. Hence a truncation (and so gauge) dependence may enter.

       However, with respect to the lattice these methods also offer 
       some advantages: there are no discretization nor volume effects, changing the 
       quark masses is simple, there is no sign problem targeting the QCD 
       phase diagram and also calculations are not a priori restricted to 
       Euclidean space. They would thus allow us for a look at
       QCD that complements the lattice, if the truncations were under control.
       It is hard to quantify their impact or to 
       systematically improve them without external input. But this is where 
       the lattice can help since it allows us to provide that input 
       nonperturbatively and untruncated.

 \item Besides this \emph{``lattice service work''} for other scientific 
       communities, calculating these $n$-point functions also provides access
       to the strong coupling and quark masses, i.e., the fundamental parameters of QCD.
       These can be obtained, for example, from the high-momentum dependence of 
       quark, gluon and ghost propagators (see, e.g., 
       \cite{Boucaud:2008gn,vonSmekal:2009ae,Burger:2012ti,Sternbeck:2012qs,Blossier:2012ef})
       or higher $n$-point functions. Also the chiral condensate 
       $\langle\bar{\psi}\psi\rangle$ can be determined in the massless limit
       (e.g., \cite{Burger:2012ti,
       Wang:2016lsv}). Furthermore, gauge-variant $n$-point functions are
       essential for the nonperturbative renormalization of hadron physics
       observables (e.g., \cite{Gockeler:2010yr}) using RI-MOM schemes 
       \cite{Martinelli:1994ty}.
       
       Since $n$-point functions are the fundamental building blocks of a 
       quantum field theory, for QCD they should show signatures of confinement and 
       chiral symmetry breaking. The chiral condensate is one example, another
       is the enhancement of the quark mass function one typically sees
       for the quark propagator in Landau and Coulomb gauges towards zero 
       momentum, i.e.,  how at low energies a quark becomes a constituent quark.
       Also confinement should be reflected, and different confinement
       scenarios connect confinement with a particular behavior of $n$-point 
       functions. In the 70's, for example, a $1/p^4$
       dependence was expected for the gluon propagator to account for a 
       linear rising quark--anti-quark potential. In contrast, other confinement
       criteria, namely the Kugo-Ojima scenario and the Gribov-Zwanziger 
       horizon condition, favor an infrared-vanishing gluon propagator
       and an infrared-diverging ghost dressing function. (We will see below,
       none of these expectations have been confirmed in lattice simulations, 
       both these objects were found being finite at $p^2=0$). So the study of 
       gauge-variant $n$-point functions gives insight in how confinement and 
       chiral symmetry breaking work in QCD. Sure these functions depend on the gauge but 
       confinement and chiral symmetry should not. There is however no 
       contradiction here, because only the descriptions differ. There is a 
       nice analogy from mechanics: The path through space of a particle in a external field may 
       look different in different coordinate systems, even though they describe
       the same path. 
\end{enumerate}

\section{Infrared behavior of the gluon and ghost propagators in Landau gauge}

Michael and I got interested in lattice studies of gauge-variant $n$-point functions around 2003 when Ernst-Michael 
Ilgenfritz brought the idea from his visit in T\"ubingen to study low-momentum 
behavior of the SU(3) gluon and ghost propagators. In T\"ubingen people were
performing similar calculations of these propagators for the SU(2) theory (later 
in collaboration with Cucchieri and Mendes) \cite{Langfeld:2001cz,Bloch:2003sk}.

A motivation for this came from new developments for the treatment of the coupled
gluon and ghost DSEs by von Smekal, Alkofer and collaborators 
\cite{vonSmekal:1997ohs,Alkofer:2000wg}. Contrarily to former expectations, they
found that the gluon propagator does not diverge in the infrared-momentum limit 
but the ghost propagator should instead diverges stronger than $1/p^2$. In fact, a 
power-law behavior for the gluon and ghost dressing functions,
\begin{equation}
 Z(p^2) \propto (p^2)^{2\kappa}\quad\text{and}\quad J(p^2)\propto (p^2)^{-\kappa}
 \qquad\text{for}\;\, p\to 0
 \label{eq:gluon_ghost}
\end{equation}
was proposed with a critical exponent $\kappa\simeq 0.595$ \cite{Zwanziger:2001kw,Lerche:2002ep}. These dressing functions
parametrize the nonperturbative momentum dependence of gluon and ghost 
propagators in Landau gauge
\begin{equation}
   D_{\mu\nu}(p^2) = \left(\delta_{\mu\nu}-\frac{p_\mu p_\nu}{p^2}\right)\frac{Z(p^2)}{p^2}\,,
   \qquad G(p^2) = \frac{J(p^2)}{p^2}
\end{equation}
and if this power-law behavior (for $p\to 0$) was true it would nicely fit to the 
Kugo-Ojima confinement criterion and the Gribov-Zwanziger horizon condition. 
Furthermore, the strong coupling constant (today aka ''Minimal MOM`` or 
''ghost-gluon coupling`` \cite{Boucaud:2008gn,vonSmekal:2009ae}) 
should settle at a finite value 
\begin{equation}
 \label{eq:alphas}
 \alpha_s(p^2) = \frac{g_0^2}{4\pi} Z(p^2) J^2(p^2) \quad\xrightarrow{p^2\to 0}
 \quad \alpha_c > 0\,.
\end{equation}

The first SU(2) lattice results \cite{Langfeld:2001cz,Bloch:2003sk} were not in 
contradiction to this. With our SU(3) lattice calculations we could however reach
smaller momenta and clearly saw this coupling to decrease \cite{Sternbeck:2005tk}, 
in disagreement to the anticipated power-law behavior. A similar trend
was found before by Furui and Nakajima \cite{Furui:2003jr}, but the error bars 
of their data still allowed to interpret this as a statistical artifact.

\begin{figure}
 \includegraphics[height=5.3cm]{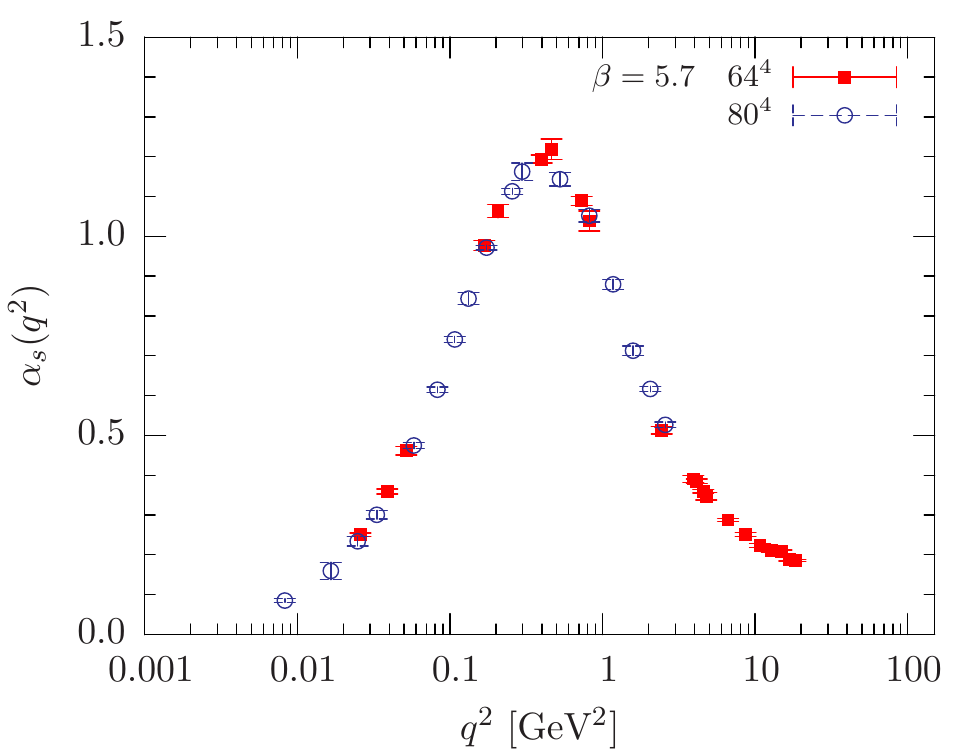}\quad
 \includegraphics[height=5.3cm]{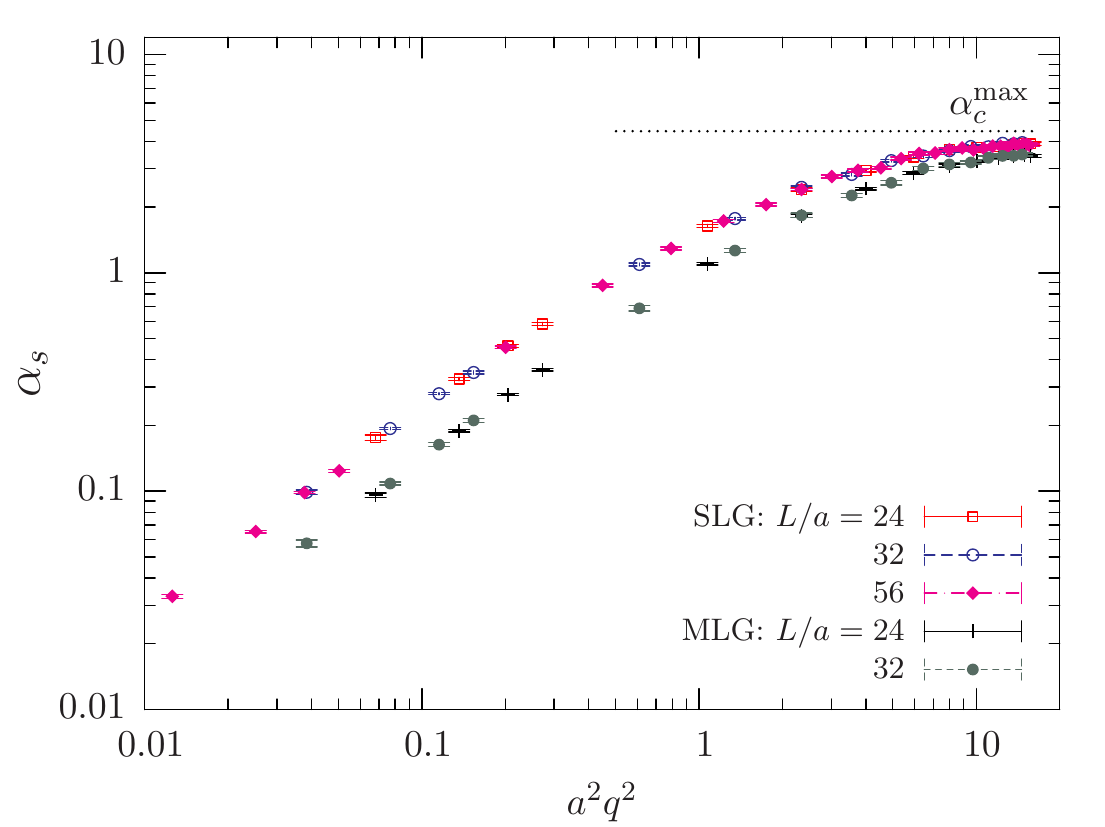}
 \caption{Left from \cite{Bogolubsky:2009dc}: $\alpha_s$ as defined in 
   Eq.\,\eqref{eq:alphas}
  versus $q^2$; Right from \cite{Sternbeck:2008mv}: 
  $\alpha_s$ in the strong-coupling limit. If there was a 
  power-law behavior the data points would collapse at $\alpha_s\simeq\alpha^{\max}_c$ for all 
  lattice momenta $a^2q^2$. The data points (SLG vs. MLG) refer to different
  discretizations of the gluon and ghost fields.}
 \label{fig:alphas}
\end{figure}

At that time it was not clear if this disagreement at low $p^2$ simply results
from a restricted lattice size. Therefore, calculations for large lattice 
sizes [up to $128^4$ for SU(2) and $96^4$ for SU(3)] were launched, 
in particular of the numerically less demanding gluon propagator. 
These efforts culminated in 2007 at the Lattice Conference in Regensburg where
the disagreement between lattice and functional methods was confirmed: both the 
SU(2) and SU(3) gluon propagator were found to approach a 
finite value at zero momentum \cite{Bogolubsky:2007ud,Cucchieri:2007md,Sternbeck:2007ug}
(see also Fig.\,\ref{fig:gl_qq_beta5p7}). 
The ghost propagator was less diverging than expected and the strong coupling 
constant found to decrease towards $p^2\to 0$ (see also Fig.\,\ref{fig:alphas}).

\medskip
Besides those efforts on the lattice, different groups also worked on 
improvements on the functional method side. Some groups even did not see any 
disagreement, because their solution of the gluon and ghost DSEs qualitatively 
agrees with the lattice results at small $p^2$ (see, e.g., 
\cite{Boucaud:2008ky,Aguilar:2008xm,Dudal:2007cw}). Another study of both
DSE and FRG equations even found there is a whole family of possible forms for
the gluon and ghost propagator's low-momentum behavior, which can be 
fixed by an additional condition on the ghost dressing function $J(0)$ 
at zero momentum \cite{Fischer:2008uz}.
In their language the afore-mentioned power-law (or conformal) behavior is the 
\emph{scaling solution} (which one gets when setting $J^{-1}(0)=0$), while for 
setting $J^{-1}(0)$ to any finite value one gets one of many 
\emph{decoupling solutions}, i.e., a gluon propagator and ghost dressing function
which are finite at $p^2=0$ and a coupling which decreases 
(as seen on the lattice). 

This additional condition on the ghost dressing function seems to cure 
the Gribov ambiguity in the Landau gauge condition; it definitely triggered new 
lattice investigations. Studies of the impact of the Gribov ambiguity were 
not new (Michael and others had performed several before), but now there was a 
definite Gribov-copy dependence that should be seen for the gluon and ghost
propagators at \hbox{low $p^2$} \cite{Fischer:2008uz}. One study was performed 
by Maas \cite{Maas:2009se}, in which the
condition on $J(0)$ was mimicked by selecting Gribov copies which at some 
finite $p^2$ give large or small values for $J(p^2)$.\footnote{Note that $J(0)$ 
is not accessible on a periodic lattice due to the trivial zero modes of the 
Faddeev-Popov operator.} Another lattice study was performed by Michael and 
myself \cite{Sternbeck:2012mf}, in which we selected Gribov copies based on 
the lowest-lying (non-zero) eigenvalue $\lambda_1$ of the Faddeev-Popov operator.

\begin{floatingfigure}{5.4cm}
 \vspace{3ex}
 \hspace{-2em}
 \includegraphics[width=5.4cm]{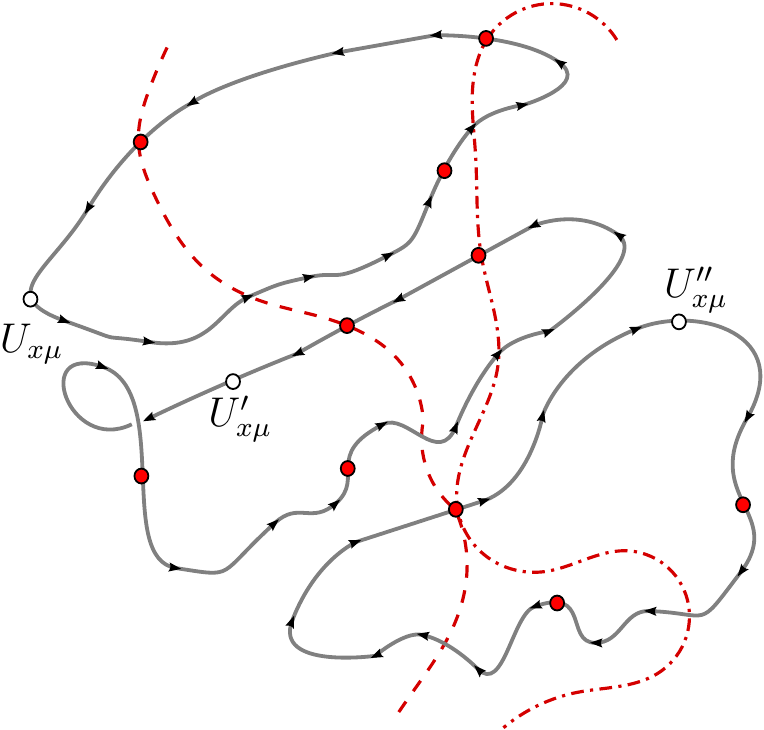}
 \caption{Artist's rendering of three gauge orbits with depiction of Gribov
  copies (red circles). A dashed and a dashed-dotted line represent a 
  particular selection of copies, one from each orbit.
  Such selection is either random or according to a particular criterion 
  (examples are given in the text).
  }
\end{floatingfigure}
To better understand how certain copies are selected, it is instructive to remember how the 
Landau gauge is implemented on the lattice. It is usually a two-step procedure, 
where in a first step gauge configurations are generated by some standard 
Monte-Carlo algorithm. In a second step, these (unfixed) configurations of 
gauge links $\{U_{x\mu}\}$ are then iteratively gauge-transformed
\begin{equation}
 \label{eq:gaugetrafo}
 U^g_{\mu}(x) = g(x) U_{\mu}(x)\, g^\dagger(x+\mu)
\end{equation}
along the gauge orbit of $U_{\mu}(x)$ until they satisfy the lattice Landau 
gauge condition 
\begin{align} \nonumber
   0 &=  \sum_{\mu} A_\mu^a(x+\mu/2) - A_\mu^a(x-\mu/2) \\ &\equiv \sum_{\pm\mu} \Im\Tr T^a U^g_{\mu}(x)
\end{align}
to machine precision. This can be achieved, for instance, by a numerical 
maximization of the lattice Landau gauge functional ($U$ is kept fixed, $g$ is varied)
\begin{equation}
 F_U[g] = \frac{1}{4V} \sum_{x,\mu} \Re\Tr U_{\mu}^g(x)\;.
\end{equation}

\noindent Also saddle points of $F_U[g]$ would satisfy the gauge 
condition. These however one does not find this way. A maximization, or 
alternatively an (accelerated) steepest-descent algorithm, will only find one 
of the many maxima of $F_U[g]$. These constitute a subset of the numerous 
Gribov copies $U^g_{\mu}(x)$, $U^{g^\prime}_{\mu}\!(x)$, 
$U^{g^{\prime\prime}}_{\mu}\!\!(x), \ldots$, which all satisfy the Landau gauge 
condition but are related by a finite gauge transformation. Choosing copies which
are global maxima of $F_U[g]$ would for example cure the Gribov problem as argued by
Zwanziger \cite{Zwanziger:1998ez}. This however is numerically hard to achieve. 
What one can achieve, and was intensively tested
by Michael and collaborators, is the method of simulated annealing which usually
finds maxima (\emph{best copies}) which are larger than those one finds using a standard 
gauge-fixing algorithm. It is unclear though if \emph{best copies} are in any
way 'better' than others.

\begin{figure*}
  \centering
  \begin{minipage}[b]{0.44\textwidth}
   \includegraphics[width=1.05\textwidth]{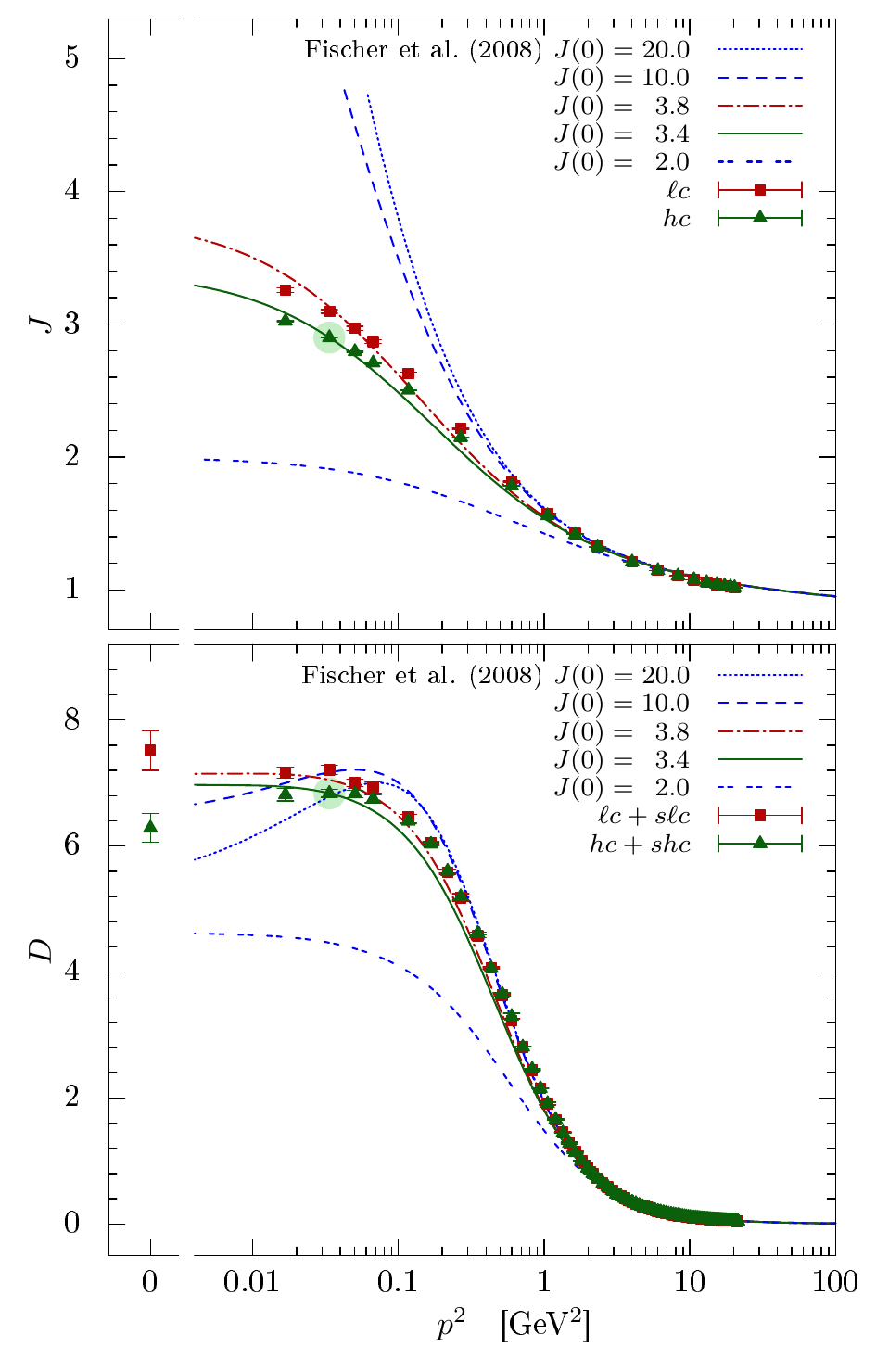}
    \caption{From \cite{Sternbeck:2012mf}: The SU(2) ghost dressing function 
     (top) and the gluon propagator (bottom) versus momentum squared. Shown 
     are our lattice data for two different selections of Gribov copies, lines 
     refer to the DSE solutions of \cite{Fischer:2008uz}.}
     \label{fig:gl_ghdress_qq}  
  \end{minipage}
  \hfill
  \begin{minipage}[b]{0.52\textwidth}
    \includegraphics[width=0.91\textwidth]{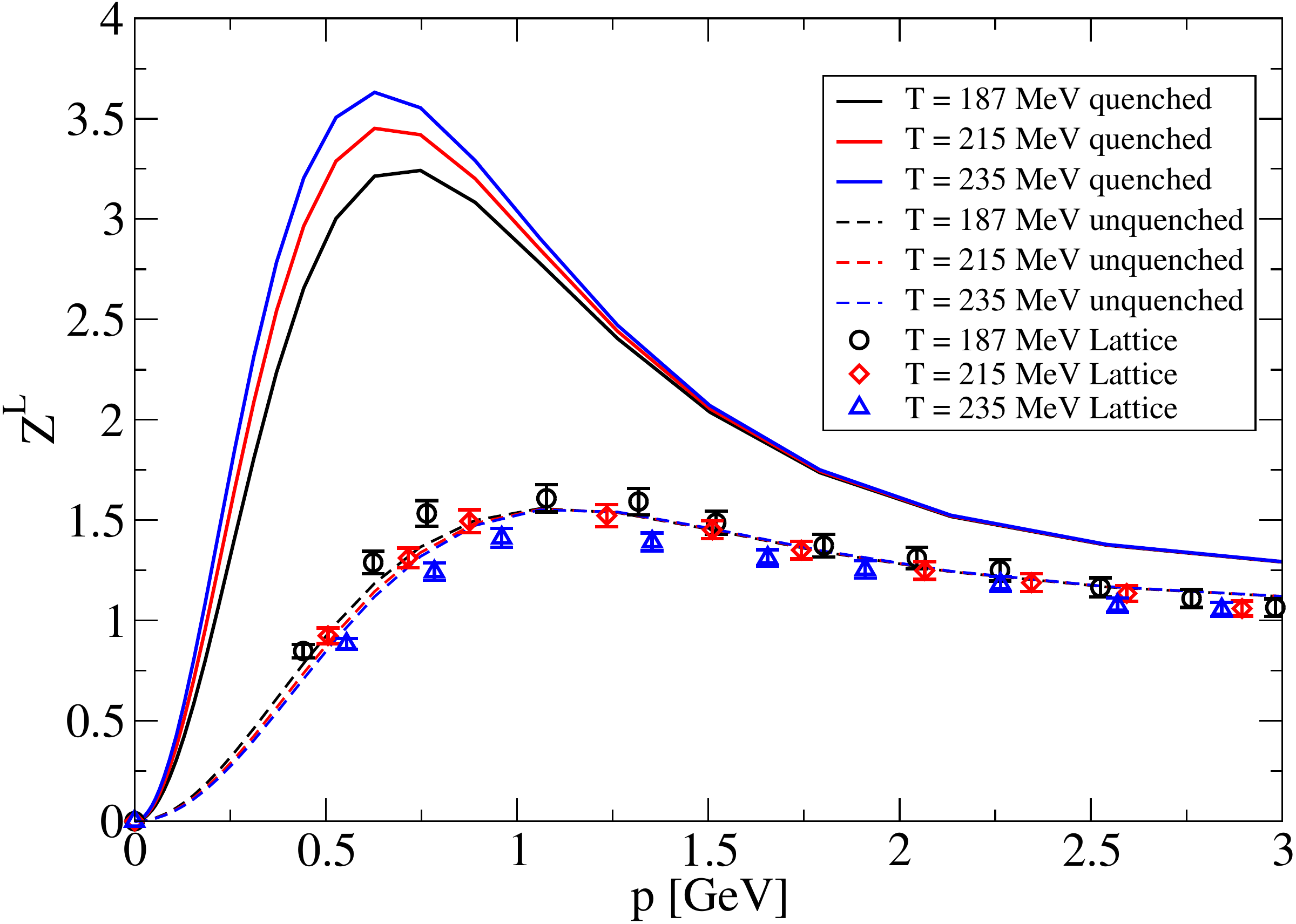}
    \includegraphics[width=0.91\textwidth]{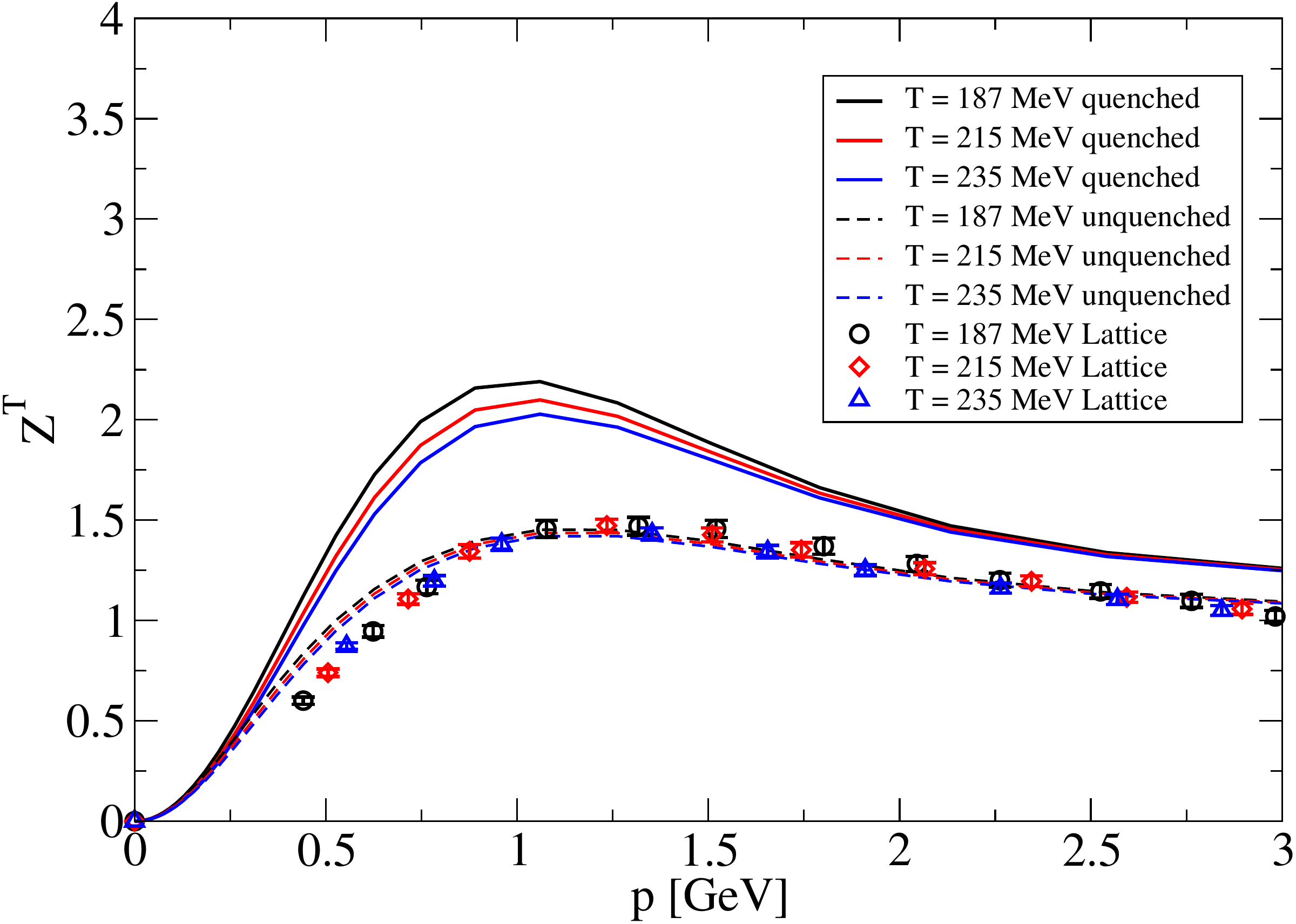}
    \caption{From \cite{Fischer:2014ata}: Longitudinal ($Z^L$, top) and 
      transversal ($Z^T$, bottom) gluon dressing function vs.\ $p$. Solid 
      lines are DSE solutions tuned such that they match our quenched 
      lattice results (not shown) from \cite{Aouane:2011fv}. Dashed lines are 
      DSE solutions which incorporate quark effects. Open symbols 
      are our $N_f=2$ data from \cite{Aouane:2012bk}.}
    \label{fig:verglMMP_ZTZL}
  \end{minipage}
\end{figure*}

Resuming to the discussion before, in \cite{Maas:2009se} Maas selected for 
each gauge configuration $\{U_{\mu}(x)\}$ that copy $\{U^g_{\mu}(x)\}$ which 
results in a large (or small) vacuum expectation value for $J(p^2)$, irrespective of the
value for $F_U[g]$. We
\cite{Sternbeck:2012mf} chose those copies with the highest ($hc$) or lowest ($lc$) value 
for $\lambda_1$ among 200 copies $\{U^g_{\mu}(x)\}$ for each configuration 
$\{U_{\mu}(x)\}$ (note, $hc$ and $lc$ refer to the entries in the legend of Fig.\,\ref{fig:gl_ghdress_qq}).
Both approaches are somehow complementary, because Gribov copies with small $\lambda_1$
give large $J(p^2)$. It was interesting to see that on 
the lattice one sees this family of decoupling solutions, albeit (for some 
yet unknown reason) only a small subset and also no scaling behavior is seen.
In our study we went even further and did a one-to-one comparison of our 
lattice data with the (decoupling) DSE results of \cite{Fischer:2008uz}. 
We did this comparison separately for two data sets 
($hc$ and $lc$) and found for both an approximate matching of lattice and DSE results 
(see Fig.\,\ref{fig:gl_ghdress_qq}).
\footnote{There is an interesting story to that: We first thought
our lattice results contradict the DSE results, since our 
gluon propagator data at small $p^2$ moves up if the ghost dressing function
gets a stronger momentum dependence for $p^2\to0$. According to \cite{Fischer:2008uz} the 
gluon propagator should decrease instead. It turned out, however, 
below a certain value for $J(0)$ the DSE results show the same (opposite) trend and also 
approximately match our data, but those solutions were simply not shown in 
\cite{Fischer:2008uz}.}

\section{The gluon and ghost propagators at finite temperature}

Studying the low-momentum behavior of the gluon and ghost propagators, to 
some extent, tries to answers an academic question, because for actual hadron physics 
or QCD thermodynamics calculations with either of the two functional methods, 
the exact form of low-momentum behavior is not that important. The interesting 
nonperturbative region is in the mid-momentum regime, i.e., at about 
0.5\ldots 3\,GeV, roughly there where the gluon dressing function shows its 
characteristic hump (Landau gauge) and where also other gauge-variant $n$-point
functions behave nonperturbatively.

In recent years there have been attempts to address the QCD phase diagram
with functional methods (DSEs and FRGEs, see, e.g., 
\cite{Fischer:2012vc,Fischer:2013eca,Fister:2013bh,Fischer:2014ata}).
As mentioned above, these methods do not suffer from the sign problem as
lattice QCD and they can thus address QCD at finite temperature and finite chemical 
potential more easily. However, these methods have to deal with the fact 
that the system of equations has to be truncated to perform calculations. 
These truncations have been improved over the years but without external input
it is hard to quantify the effect on the final result.

Together with a PhD student (Aouane) of Michael we have studied the gluon 
and ghost propagator in Landau gauge at finite temperature. One study was for quenched
QCD another for QCD with $N_f=2$ degenerate (twisted-mass) Wilson fermions.
This latter project built upon the vast set of gauge field configurations from the tmfT 
collaboration. Our results were published in 2011 and 2012 
\cite{Aouane:2011fv,Aouane:2012bk} and immediately served as input and benchmark
for DSE/FRGE-based studies of the QCD phase diagram (e.g., 
\cite{Fischer:2013eca,Fischer:2014ata}). 

An example is shown in Fig.\,\ref{fig:verglMMP_ZTZL} which has been taken from 
the DSE-based study of the QCD phase diagram in \cite{Fischer:2014ata}. In this
figure, the DSE results for the transversal and 
longitudinal gluon dressing functions
\footnote{Note at finite temperature the Landau-gauge gluon propagator is 
parametrized by two dressing functions, $Z^T$ and $Z^L$.
}
are shown for quenched QCD (solid lines) and for QCD with $N_f=2$ fermions 
(dashed lines). The DSE solutions for $N_f=0$ are tuned such that 
they agree with quenched lattice data (not shown), for example, from 
\cite{Aouane:2011fv,Silva:2013maa}. It is reassuring that then their 
DSE solutions for $N_f=2$---using the quenched DSE solutions and a 
suitable ansatz (truncation) for the fermionic back reaction---more or less 
match with our $N_f=2$ lattice data (open symbols in Fig.\,\ref{fig:verglMMP_ZTZL}).
Our lattice data thus gave strong support for the truncation used in 
\cite{Fischer:2014ata}, based on which an approximate location of the 
critical end point in the QCD phase diagram could be provided.

\section{Triple-gluon and quark-gluon vertex in Landau gauge}

\begin{floatingfigure}[r]{5cm}
\centering
\hbox{\includegraphics[width=2.3cm]{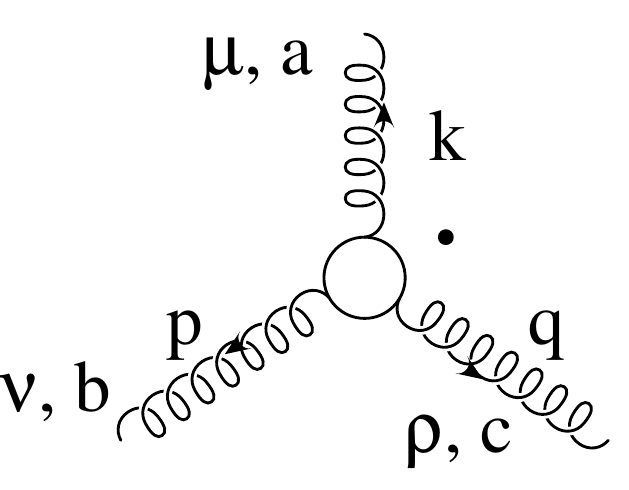}\quad
\includegraphics[width=2cm]{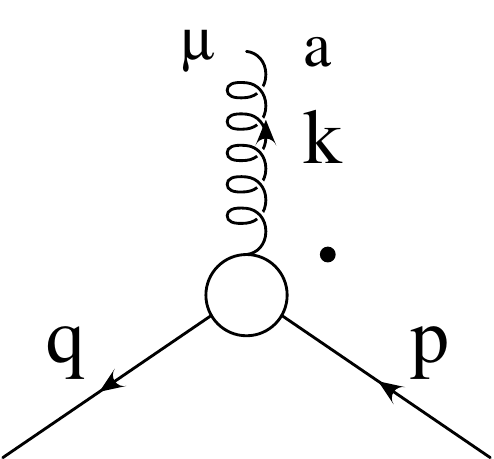}}
\caption{Tripe-gluon and quark-gluon vertex; taken from \cite{Alkofer:2000wg}.}
\label{fig:3gluon_quarkgluon}
\end{floatingfigure}
For similar efforts in the future it is important to proceed with lattice 
calculations of gauge-variant $n$-point functions. Basically for all $n$-point 
functions with $n>2$ not much is known about the nonperturbative structure and 
lattice input is much appreciated. The 3-gluon and quark-gluon vertex at 
zero and finite temperature, for example, are good next candidates. They are of 
prime importance for the study of bound-state equations, for studies of QCD 
at finite temperature and chemical potential and implicitly are always 
also needed for the developments of improved truncation schemes (see, e.g., \cite{Eichmann:2016yit} for 
truncations beyond 'Rainbow-Ladder').

Compared to 2-point functions (propagators), lattice calculations of vertex 
functions are much more involved, however. To estimate the momentum dependence
of, say the gluon propagator, one needs about 100 well-decorrelated gauge field 
configurations (if translation invariance is exploited). Moreover, a propagator 
is typically parametrized by only one or two form factors, and 
these one often gets straightforward from the Monte-Carlo estimate.
A vertex, in comparison, comes with a much richer tensor structure and 
its form factors in general depend on at least two momenta. Furthermore, to 
reliably estimate the corresponding $n$-point Green's function
a much increased statistics is needed (for our results in 
Fig.\,\ref{fig:3gluon_quarkgluon} we had to analyze 
about 1000 gauge field configurations for each parameter set). Another 
complication is the fact that it is the $n$-point (Green's) function one 
calculates on the lattice, not the vertex. The latter is extracted from 
certain combinations of the Monte-Carlo estimates for the 2- and 3-point 
functions. In fact, lattice studies give us access to the quark propagator ($S$), 
the gluon propagator ($D_{\mu\nu}$) and to the 3-point Green's functions
\begin{align}
 \label{eq:3Gluon}
 G_{\mu\nu\rho}(p,q) &= \sum_{xy}e^{-ipx-iqy} \left\langle A_\mu(x)A_\nu(y)A_{\rho}(0)\right\rangle
 \; = D_{\mu\lambda}(p)\, D_{\nu\sigma}(q)\,D_{\rho\omega}(-p-q)\,\Gamma^{AAA}_{\lambda\sigma\omega}(p,q)\,, \\
 \label{eq:QGluon}
 V_{\mu}(p,q) &= \sum_{xy}e^{-ipx-iqy}\left\langle \bar{\psi}(x)\psi(y) A_\mu(0)\right\rangle
 \hspace{1em} = S(p)\, S(q)\, D_{\mu\nu}(-p-q)\, \Gamma^{\bar{\psi}\psi A}_\nu(p,q) \,.
\end{align}
The last terms demonstrate, the vertex $\Gamma$ is obtained through an 
amputation of the corresponding propagators. It is this amputated (and often 
statistically noisy) object from which one extracts the several form factors 
of the vertex.

The 3-gluon and quark-gluon vertex are parametrized by 14 and 12 form factors, respectively,
\footnote{For $\Gamma^{\bar{\psi}\psi A}_\mu$ we quoted the ''Ball-Chiu`` 
form, according to which the vertex splits into a transverse part (second term) 
and another that satisfies the Slavnov-Taylor identity (first term).
}
\vspace{-1ex}
\begin{align}
  \Gamma^{AAA}_{\mu\nu\lambda}(p,q) &= \sum_{i=1}^{14} f_i(p,q) 
    \, P^{(i)}_{\mu\nu\lambda}(p,q) \,,\\
  \Gamma^{\bar{\psi}\psi A}_\mu(p,q) &= \sum^4_{i=1}\lambda_i(p,q)\,L^{(i)}_{\mu}(p,q)
   + \sum^8_{j=1} \tau_j(p,q)\, T^{(j)}_{\mu}(p,q)\,.
\end{align}
The $P$s, $L$s and $T$s are suitably chosen base vectors in the momenta $p$ and $q$
(see, e.g., \cite{Skullerud:2002ge,Gracey:2014mpa}) and the scalars 
$f_i$, $\lambda_i$ and $\tau_i$ are the corresponding form factors 
which contain all the nonperturbative information of the vertex. They are functions
of $p^2$, $q^2$ and $p\cdot q$ and, depending on the base and momenta
$p$ and $q$, are accessible through Eqs.\,\eqref{eq:3Gluon} and \eqref{eq:QGluon}.
From perturbation theory, the (off-shell) form factors are known up two-loop 
order (see, e.g., \cite{Gracey:2014mpa} are references therein). For special choices
of momenta even three-loop results are available \cite{Chetyrkin:2000dq}.

In Landau gauge, only the transverse part of the 3-gluon vertex is relevant, because
it is always sandwiched by transverse gluons. The transversally projected 3-gluon
vertex can be parametrized by four (transverse) form factors $F_{1,\ldots,4}$ (see, e.g.,
\cite{Eichmann:2014xya}). These form factors are accessible on the lattice, 
in the past, however, only the projection
\begin{equation}
 G_1(p,q) =
\frac{\Gamma^{(0)}_{\mu\nu\rho}}{\Gamma^{(0)}_{\mu\nu\rho}}
\frac{G_{\mu\nu\rho}(p,q,p-q)}{D_{\mu\lambda}(p)
D_{\nu\sigma}(q)D_{\rho\omega}(p-q)
\,\Gamma^{(0)}_{\lambda\sigma\omega} }
\end{equation}
of the vertex onto its tree-level form, $\Gamma^{(0)}(p,q)$, has been considered.
For SU(2) Yang-Mills theory it was found \cite{Cucchieri:2008qm} that $G_1$ shows
a zero-crossing at some small momentum $\vert p\vert= \vert q\vert$. This was
unexpected and triggered some interest. Recently, this zero-crossing has been 
confirmed by two independent lattice studies 
\cite{Athenodorou:2016oyh,Duarte:2016ieu}.

In 2014 Michael and I were granted computing time for a study of the 3-gluon and 
quark-gluon vertex. Our plan has been to study not only a particular channel 
but the full nonperturbative structure of both vertices. Since then many $N_f=2$ 
gauge field configurations (provided by the RQCD collaboration) have been fixed 
to Landau gauge and the necessary 2- and 3-point functions were calculated. 
We further generated some quenched ensembles to analyze unquenching effects and 
to reach much lower momenta. The analysis of the 3-gluon vertex was done in
collaboration with a student of mine (Balduf).
The study of the quark-gluon vertex is a collaborative effort with
colleagues from Australia (K{\i}z{\i}lers\"u, Williams), Portugal (Oliveira, Silva) 
and Irland (Skullerud). 

For both vertices we have first (preliminary) data. For example, for the 
3-gluon vertex we see that the leading (transverse) form factor is $F_1$ while 
the remaining three are much smaller in comparison and also show only a weak momentum 
dependence (we studied cases where $\vert p\vert = \vert q\vert$). That 
is, $G_1\approx F_1$ contains most of the 3-gluon vertex's 
nonperturbative information. Furthermore we have looked at the zero-crossing, 
but cannot confirm it yet (see Fig.\,\ref{fig:3gluon_quarkgluon}, left). Our 
current data for $G_1$ approaches zero but does not clearly cross it (taking 
the numerical fluctuations as such). This remains to be clarified and will
be discussed in due course \cite{Balduf2016}.

Also for the quark-gluon vertex preliminary data for the soft-gluon kinematic 
(zero gluon momentum) is available. In this kinematic only the form 
factors $\lambda_1$, $\lambda_2$ and $\lambda_3$ are non-zero. An example of the
(preliminary) data for $\lambda_1$ is shown in Fig.\,\ref{fig:3gluon_quarkgluon}, 
right; it is the same as in \cite{Oliveira:2016muq}. One sees we have to deal 
with quark mass effects and significant discretization effects at larger 
momenta but apart from that, the data at lower momenta clearly shows the quark-gluon vertex 
becomes quickly nonperturbative if $p^2$ drops below 2 or 3 GeV$^2$. Similar 
is seen for $\lambda_2$ and $\lambda_3$. Above this momentum regime, the 
quark-gluon vertex seems to share much with its tree-level form. Also this will 
be discussed in more detail in due course \cite{QGV}.

\begin{figure*}
 \centering
 \hbox{\includegraphics[height=5cm]{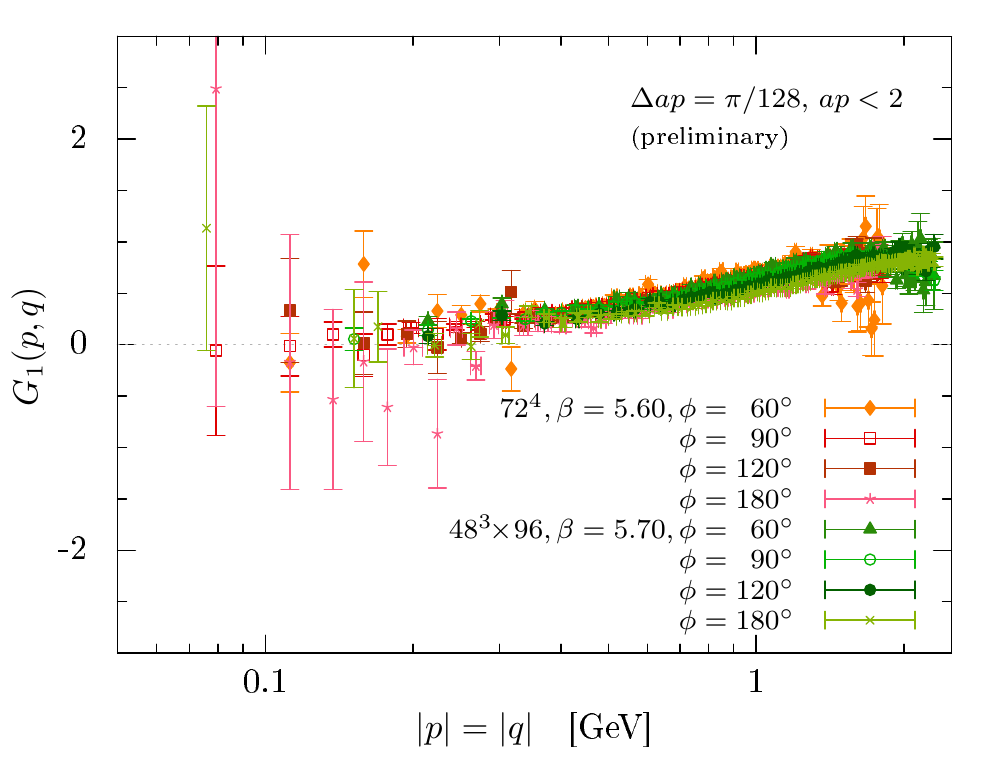}\quad
 \includegraphics[height=5cm]{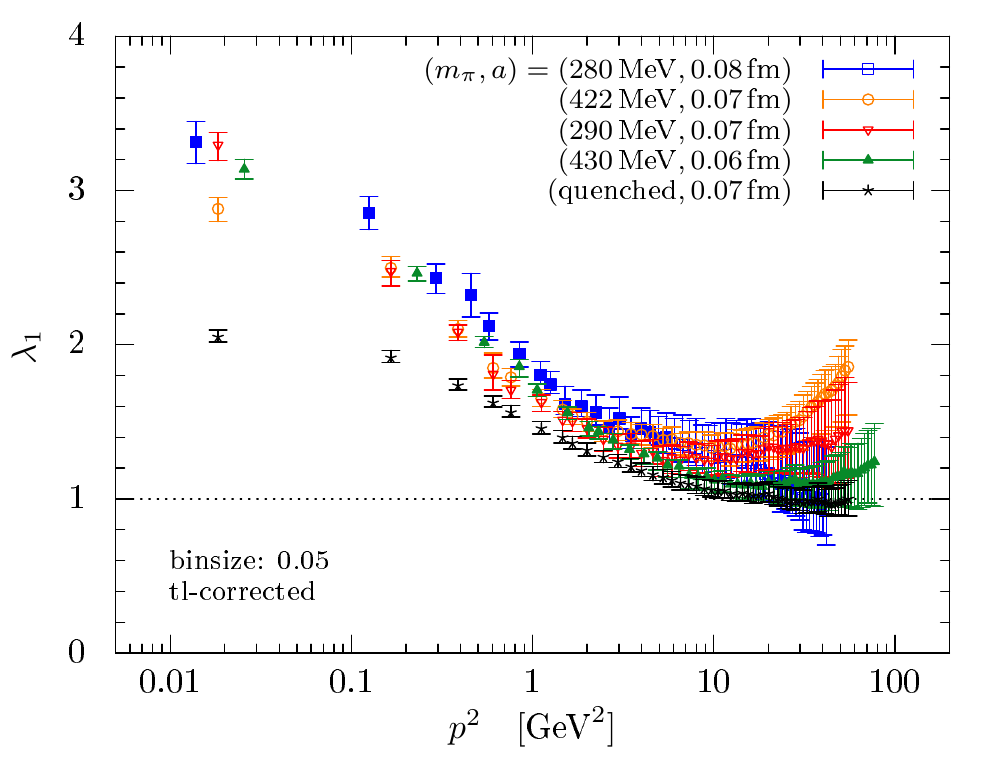}}
 \caption{Form factors versus momentum, which describe the deviation from the tree-level form
  of the 3-gluon vertex (left) and of the quark-gluon vertex (right).
  The data is yet preliminary and not renormalized. For the 3-gluon vertex we show data
  for different angles $\phi$ between $p$ and $q$. For the quark-gluon-vertex
  only the soft-gluon kinematic (gluon has zero momentum) is shown. \vspace{-3ex}}
  \label{fig:3gluon_quarkgluon}
\end{figure*}

\centerline{----------------------------------------}

Let me conclude this brief summary of our lattice studies of gauge-variant 
$n$-point functions. I feel grateful for the time spent with Michael and our 
collaborators as we traveled together on this 14 year endeavor. We will miss 
a passionate teacher, a distinguished scientist, a person of integrity 
and a faithful colleague.

\newpage
\bibliography{references}

\end{document}